\documentstyle[prl,aps,epsf,multicol]{revtex} 
\tighten
\renewcommand{\narrowtext}{\begin{multicols}{2}
\global\columnwidth20.5pc} 
\renewcommand{\widetext}{\end{multicols}
\global\columnwidth42.5pc} 
\begin{document} 
\draft 
\title{Composite fermions in a long-range random magnetic field:\\
Quantum Hall effect versus Shubnikov-de Haas oscillations}
\author{A. D. Mirlin,$^{1,*}$ D. G. Polyakov,$^{2,\dagger}$ and
P. W\"olfle$^1$}
\address{$^1$ Institut f\"ur Theorie der Kondensierten Materie,
Universit\"at Karlsruhe, 76128 Karlsruhe, Germany\\ $^2$ Institut
f\"ur Theoretische Physik, Universit\"at zu K\"oln, 50937 K\"oln,
Germany}
\maketitle
\begin{abstract} We study transport in a smooth random magnetic field,
with emphasis on composite fermions (CF) near half-filling of the
Landau level. When either the amplitude of the magnetic field
fluctuations or its mean value $\overline{B}$ is large enough, the
transport is of percolating nature. While at $\overline{B}=0$ the
percolation effects enhance the conductivity $\sigma_{xx}$, increasing
$\overline{B}$ (which corresponds to moving away from half-filling for
the CF problem) leads to a sharp falloff of $\sigma_{xx}$ and,
consequently, to the quantum localization of CFs. We demonstrate that
the localization is a crucial factor in the interplay between the
Shubnikov-de Haas and quantum Hall oscillations, and point out that
the latter are dominant in the CF metal.  
\end{abstract} 
\narrowtext

The idea of composite fermions (CF) \cite{Jain89,Halperin93,Lopez91}
has been remarkably successful in accounting for the transport
properties of strongly correlated two-dimensional electrons in high
magnetic fields, especially near half-filling $\nu = \frac{1}{2}$ of
the lowest Landau level. Within this model, each electron is replaced
by a fermion carrying two flux quanta of a fictitious magnetic field
oriented oppositely to the external field ${\bf B}_{ext}$. At
half-filling, the average of this Chern-Simons field cancels the
external one exactly, thus offering an elegant explanation of the
observed gapless state with a well-defined Fermi surface. Away from
$\nu = \frac{1}{2}$, the composite fermions experience, in the
mean-field sense, a net effective field $\overline{\bf B}=(1-2\nu){\bf
B}_{ext}$. An integer number $p$ of filled Landau levels of the
fermions then corresponds to the fractional quantum Hall (QH) plateau
at $\nu=p/(2p\pm 1)$.  This mapping of the singularities in the
strongly interacting electron system at fractional fillings onto
integer QH transitions is a particularly attractive feature of the CF
picture.

Confidence that the CF picture does indeed capture the essential
physics of the $\nu={1\over 2}$ problem is strengthened by a number of
observations of Fermi-surface features at half-filling in experiments
with acoustic waves \cite{Willett95} and in direct transport
measurements \cite{Du94}. In particular, in the vicinity of
half-filling the dissipative resistivity $\rho_{xx}^e(B_{ext})$
exhibits magnetooscillations \cite{Du94} which look very much like the
familiar Shubnikov-de Haas (SdH) effect.  However, there is an
essential difference: the oscillations near $\nu={1\over 2}$ appear to
be much more strongly damped as compared to the conventional metallic
phase. Generally speaking, this is in line with the fact that
$\rho_{xx}^e$ at $\overline{B}=0$ is orders of magnitude larger than
at $B_{ext}=0$. More specifically, the enhanced damping is understood
in terms of strong scattering on spatial fluctuations of the
Chern-Simons field. The fluctuations of the effective magnetic field
$B({\bf r})=[1-2\nu({\bf r})]B_{ext}$ around the mean value
$\overline{B}$ are generated by the inhomogeneous distribution of the
electron density and therefore the local filling factor $\nu({\bf r})$
due to screening of the impurity potential. This mechanism of
scattering dominates \cite{Halperin93} the conductivity of the CF
metal, whereas the direct scattering on the scalar random potential is
of little importance.

This Letter is concerned with the disorder-induced localization of the
composite fermions and its implications for transport measurements. In
particular, we address the question as to the nature of the
magnetooscillations near half-filling. Though it has become customary
to identify the oscillations with the SdH effect \cite{Du94}, we argue
that they are due to the quantum localization responsible for the QH
effect. Specifically, the SdH and QH oscillations coexist with each
other; what we show is that the SdH oscillations are much more
strongly damped as compared to those induced by the quantum
localization. We also demonstrate that the effective magnetic field
creates a classical percolation network, which drastically suppresses
the CF conductivity. It is this classical localization effect that
enhances the quantum localization leading to the QH oscillations.

The model we deal with is that of non-interacting fermions in a random
magnetic field (RMF) $\overline{B}+B({\bf r})$ with the average
$\overline{B}$ and the correlator $\left<B(0)B({\bf
r})\right>=B_0^2F(r)$. We wish to calculate the CF conductivity as a
function of $\overline B$. The measurable electron resistivity
$\rho_{\mu\nu}^e$ can be directly expressed in terms of the CF one
\cite{Halperin93}: $\rho_{xx}^e=\rho_{xx}$,
$\rho_{xy}^e=\rho_{xy}-2h/e^2$. Disorder is assumed to be created by
ionized impurities randomly distributed with a sheet density $n_i$
(for simplicity let it be equal to the electron density $n$) in a thin
layer separated from the electron gas by an undoped spacer of width
$d\gg n^{-1/2}$. The correlation function of the RMF is then
parameterized by $B_0=k_F/\sqrt{2}d$ and $F(r)=(1+r^2/4d^2)^{-3/2}$,
where $k_F^2=4\pi n$ and $\hbar=c=e=1$.

The RMF is characterized by two length scales: the correlation radius
$d$ and the cyclotron radius in the field $B_0$, $R_0 = k_F/B_0$.
Defining the parameter $\alpha = d/R_0$, we can distinguish the
weak-RMF regime $\alpha \ll 1$, where the mean free path $l\gg R_0\gg
d$, and the regime of strong fluctuations $\alpha\gg 1$, where one
should expect drastic deviations from the Drude picture. Though
$\alpha=1/\sqrt{2}$ for the CFs at half-filling, it is instructive to
get a feeling for the problem by exploring these two limiting cases.
Also, in principle one can change $\alpha$ by varying the ratio of the
impurity and electron densities, since $\alpha=(n_i/2n)^{1/2}$.

We start with the limit of weak $\overline B$ where the quasiclassical
treatment is accurate \cite{Aronov95}. At $\alpha \ll 1$, CF
trajectories are only slightly bent on the scale of $d$, so that the
Born approximation \cite{Halperin93,Aronov95} is valid. Accordingly,
for the transport scattering time one gets $1/\tau_{tr} =
(B_0^2/mk_F)\int_0^\infty dr F(r) = 2\alpha^2v_F/d$, where the CF
effective mass $m=k_F/v_F$ is introduced. The conductivity at zero
$\overline B$ then reads (in units of $e^2/h$) \begin{equation}
\label{e1} \sigma_{xx}=k_Fd/4\alpha^2,\quad\alpha\ll 1~.
\end{equation} If the size of inhomogeneities had been strictly zero
(``white noise" disorder), one would have obtained at the
quasiclassical level the magnetoresistance
$\Delta\rho_{xx}/\rho_{xx}=0$. To get a finite magnetoresistance, one
has to include the effect of the field $\overline B$ on the
``ballistic" scale $d$. To this end we write the equation of motion in
the form $d\phi/dt=\omega_c+B[{\bf r}(t)]/m$, where $\phi$ is the
angle of the velocity on the Fermi surface and $\omega_c={\overline
B}/m$. The velocity correlation function $v_F^2\left<\cos
[\phi(t)-\phi(0)]\right>$ is then given by
$v_F^2\left<\cos\left({1\over m}\int_{0}^{t}\!dt_1B[{\bf
r}(t_1)]\right) \right>$. Here $\left<\ldots\right>$ denotes an
ensemble average over exact classical trajectories ${\bf r}(t)$. In
the limit of small $\alpha$, the classical dynamics of the system is
fully chaotic (uncorrelated diffusion) and so $B[{\bf r}(t)]$ can be
treated as a Gaussian variable with respect to the disorder
averaging. Accordingly, the conductivity is \begin{eqnarray}
\sigma_{xx}+i\sigma_{xy}&=&{k_F^2\over 2m}\int_0^\infty\!\!dt
e^{i\omega_ct-S(t)}~, \nonumber\\ S(t)&=&{1\over
2m^2}\int_0^t\!\!dt_1\!\!\int_0^t\!\!dt_2\left< B[{\bf r}(t_1)]B[{\bf
r}(t_2)]\right>~. \end{eqnarray} The double integral can be rewritten
to give $S(t)=(B_0/m)^2\int_0^{t}\!dt'(t-t')\left<F(|{\bf r}(t')-{\bf
r}(0)|)\right>$. The argument of $F$ is readily expanded at
$t\ll\tau_{tr}$ in powers of $\omega_c$: \begin{equation} |{\bf
r}(t)-{\bf r}(0)|\simeq v_Ft[1-(\omega_ct)^2/24]~.  \end{equation} We
can identify two different contributions to the magnetoresistance. One
comes from a $\overline B$ dependent correction to
$\tau_{tr}^{-1}=\lim_{t\to\infty}t^{-1}S(t)$. Substituting Eq.\ (3)
into $S(t)$ we get $\tau_{tr}^{-1}({\overline
B})\simeq\tau_{tr}^{-1}(0)[1+ (d^2/2R_c^2)\ln (l/d)]$, where
$R_c=v_F/\omega_c$ is the Larmor radius in the field $\overline
B$. This correction would yield a positive
$\Delta\rho_{xx}/\rho_{xx}$. Next, we should take into account that
the relaxation kernel is not exactly the simple exponential
$e^{-t/\tau_{tr}}$; specifically, $S(t)\propto t^2$ at $t\alt
d/v_F$. This gives a 4 times larger negative contribution to the
magnetoresistance, so that in total \begin{equation}\label{e4}
{\Delta\rho_{xx}\over\rho_{xx}}=-{3\over 2}{d^2\over R_c^2}\ln {l\over
d}~. \end{equation} This result agrees with that derived in Ref.\ 7 by
a different method. The sign of $\Delta\rho_{xx}/\rho_{xx}$ found for
$\alpha \ll 1$ is opposite to the one observed experimentally near
$\nu={1\over 2}$. The discrepancy is likely due to the fact that the
condition of weak RMF, $l\gg d$, is not met in the experiments.

Let us now turn to the strong-RMF regime, $\alpha\gg 1$, keeping
$\overline{B}=0$. The seemingly innocent assumption about the chaotic
character of the particle dynamics, which enabled us to represent the
conductivity in the form (2), is not valid any more. Most particles
are now out of play since they are caught in cyclotron orbits drifting
along the closed lines of constant $B({\bf r})$ (``van Alfv\'en
drift"). In the adiabatic limit, their drift trajectories are periodic
and so do not contribute to the conductivity. Still, however large
$B_0$ is, there are classical paths which are not localized and
percolate through the system by meandering around the lines of zero
$B({\bf r})$. The conductivity is determined by the particles that
move along these extended ``snake states" \cite{Chklovskii93}. Note
that there is one single percolating path on the manifold of the
$B({\bf r})=0$ contours; yet, the conductivity is nonzero since the
snake-state trajectories form a bundle of finite width, $R_s\sim
d/\alpha^{1/2}$. The conducting network is made up of those
snake-states that can crossover from one critical zero-$B$ line of
length $L_s\sim \alpha^{7/6}d$ (for a review of the percolation theory
see Ref.\ 9) to another. The coupling of two adjacent percolating
clusters occurs near the critical saddle-points of $B({\bf r})$, which
are nodes of the transport network. The characteristic distance
between the nodes, i.e. the size of the elementary cell $\xi_s$, is
then $\alpha^{2/3}d$ \cite{Isichenko92}. On length scales longer than
$\xi_s$, the particle dynamics can be viewed as fully stochastic. We
estimate the macroscopic diffusion coefficient as $D\sim\nu_sD_s$,
where $\nu_s\sim L_sR_s/\xi_s^2$ is the fraction of particles residing
in the delocalized snake-states, $D_s\sim\xi_s^2v_F/L_s$ their
diffusion coefficient. We thus have $D\sim v_FR_s$ and,
correspondingly, \begin{equation} \sigma_{xx}\sim
k_Fd/\alpha^{1/2},\quad\alpha\agt 1~. \end{equation} It is worth
noting that by comparison with the Born approximation [Eq.\ (1)], the
conductivity is $\sim\alpha^{3/2}$ times larger (though the
localization effects are strong and naively one might have expected
the opposite). This is consistent with the experimental observation
\cite{Du94} that the CF conductivity at half-filling is a factor of
$\sim 5$ larger than the perturbative-in-$\alpha$ result
\cite{Halperin93} $\sigma_{xx}=k_Fd/2$. Let us also note that
$\sigma_{xx}$ given by Eq.\ 5 is larger by a factor of
$\sim\alpha^{1/2}$ than that obtained for $\alpha\gg 1$ in
\cite{Khveshchenko96} by using an ``eikonal approach''. The fault in
\cite{Khveshchenko96} is not with the quasiclassical approximation
itself, but with the method of disorder averaging, which is incorrect
in principle at large $\alpha$; in particular, it neglects the
localization of particles and the percolating character of the
transport through the snake states. Whether the experimentally
observed positive magnetoresistance can be understood within the
percolation picture is not clear at present.

We now consider the regime of strong $\overline B$, which is defined
for $\alpha\sim 1$ by the condition $\overline{B}\gg B_0$, or
equivalently, $R_c \ll d$. Since $d/R_c=k_Fd/2p$, where $p$ is the
number of filled CF Landau levels, and experimentally $k_Fd\sim 10\div
15$, the range of $p$ where the magnetooscillations are observed,
$p\alt 7$, undoubtedly requires a strong-$\overline B$ treatment.  At
$\overline{B}\gg B_0$, the CF dynamics is a slow van-Alfv\'en drift of
the cyclotron orbits along the lines of constant $B({\bf r})$, i.e.\
the conductivity is again determined by a percolation network of
trajectories close to the $B({\bf r})=0$ lines. Naively one may well
think the percolation picture is very much like the one we dealt with
at zero $\overline B$ and $\alpha\gg 1$. In actual fact, there is a
crucial difference. Specifically, now there is no stochastic mixing at
the nodes of the percolation network: unlike the snake states at
$\overline{B}=0$, the rapidly rotating cyclotron orbits pass
harmlessly through the critical saddle-points of $B({\bf r})$ without
changing to the adjacent cell. In the high-$\overline B$ limit, the
mixing occurs on the links of the network and is only due to the weak
scattering between the drift trajectories.

In order to calculate the conductivity at $\overline{B}\gg B_0$, we
first need to integrate out the fast cyclotron rotation, taking care
not to lose the effect of the non-adiabatic mixing.  A similar
question for the electron system in a random scalar potential was
recently addressed in \cite{Fogler97}. We write the equation of
motion in the form \begin{equation}
z(t)=iv_F\int_0^{t}dt'e^{i\int_0^{t'}dt''\Omega [z(t'')]}~,
\end{equation} where $z(t)=x+iy$ stands for the exact trajectory ${\bf
r}(t)=(x,y)$ in the field $\overline{B}+B(z)$, and $\Omega
(z)=\omega_c+B(z)/m$. It is convenient to introduce the guiding center
coordinate $\zeta (t)=z(t)-R_c[\zeta (t)] e^{i\varphi(t)}$, where
$R_c$ is the local cyclotron radius at the point $\zeta (t)$ of the
drift trajectory and $\varphi (t)=\int_0^{t} dt'\Omega_c[\zeta (t')]$
is expressed in terms of the local cyclotron frequency.  The velocity
of the guiding center is given by $\dot\zeta (t) = iv_F\Big( e^{i
\int_0^t dt'\Omega [z(t')]} - e^{i\varphi (t)}\Big)-\dot
R_ce^{i\varphi}$.  After averaging over the cyclotron motion (denoted
by $\langle\ldots\rangle_c$), the first term leads to the usual
expression for the drift velocity ${\bf v}_d = \langle \dot\zeta
\rangle_c = \Big[mv_F^2/2(\overline{B} + B)^3\Big]\Big(\nabla B \times
(\overline{\bf B} + {\bf B})\Big)$.  At this level the drift occurs
strictly along the lines of constant $B({\bf r})$, so that the
conductivity would be zero.  The time dependence of the cyclotron
radius in the second term gives rise to the leading non-adiabatic
contribution $\delta {\bf v}_d = \langle(2 v_d^2/v_F)e^{i
\varphi}\rangle_c$, allowing the particle to perform a random walk
around the $B({\bf r})=0$ contour. Because of the rapidly oscillating
factor $e^{i\varphi}$, $\delta {\bf v}_d$ will be exponentially
small. We define the diffusion coefficient associated with the motion
perpendicular to the drift trajectory as $D_\bot = \frac{1}{4}
\lim_{t\rightarrow \infty} t^{-1} \langle[\int \delta{\bf v}_d
(t)dt]^2\rangle_d \propto e^{-W}$, where $\langle\ldots\rangle_d$
denotes the average over the impurity ensemble along the drift lines.

Expressing the phase $\varphi$ by integration along the contour,
$\varphi = \omega_c\int d l v_d^{-1}(l)$, and shifting the integration
into the complex plane one observes that the exponent $W$ is
determined by the phase $\varphi$ picked up at the singular point of
the correlator $\langle B(0)B(r)\rangle$ at $r=2id$. Consequently,
\begin{equation} W=-\ln\left<e^{i\omega_c\int_0^{2id}dlv_d^{-1}(l)}
\right>~, \end{equation} where the integration should be done along
the straight line connecting the points $l=0$ and $l=2id$. It follows
that the dominant contribution to $D_\bot$ comes from rare
fluctuations in which the drift velocity greatly exceeds the typical
value $v_F(B_0/\overline{B})^2$. Indeed, $W$ can be written as a sum
of two terms, \begin{equation} W_1={1\over 2}\int\!\!{d^2{\bf k}\over
(2\pi)^2}{|v^0_{dx_{\bf k}}|^2\over \left<v_{dx}v_{dx}\right>_{\bf
k}}~,\,W_2=i\omega_c\int_0^{2id}\!\!\!\!\!{dx\over
v_{dx}^0(x,0)}~. \end{equation} Here $W_1$ represents the probability
for the optimum fluctuation $v^0_{dx}({\bf r})$ to occur, $W_2$ stands
for the probability of the non-adiabatic scattering on this
fluctuation. Solution of the variational equation $\delta W/\delta
v^0_{dx}=0$ then yields $v^0_{dx}({\bf r})=v_F(B_0/\overline{B}){\cal
F}({\bf r}/d)$, where $\cal F$ is a dimensionless function.  
We thus get \begin{equation} W=c(\overline{B}/B_0)^2~, \end{equation}
where the numerical coefficient $c\sim 1$. Choosing $v_{dx}^0({\bf
r})\propto \left<v_{dx}(0)v_{dx}({\bf r})\right>$ as a trial function
with the variational parameter $v_{dx}^0(0)$, we obtain the estimate
$c\simeq 1.6$.

The non-adiabatic mixing of drift trajectories yields the conductivity
$\sigma_{xx}\sim k_F\delta$, where $\delta$ is an effective width of
the links of the percolation network. This $\delta$ obeys the equation
$\delta^2\sim D_\bot L(\delta)/v_d$, which is the condition of
connectivity of the network. Here $L(\delta)\sim d(d/\delta)^{7/3}$ is
the characteristic perimeter of the cells
\cite{Isichenko92}. Consequently, \begin{equation}
\sigma_{xx}=k_Fd\times f({\overline{B}\over B_0})\exp \left[-{3c\over
13}\left({\overline{B}\over B_0}\right)^2\right]~, \end{equation}
where $f(x)$ is a power-law function. The conductivity is seen to fall
off sharply beyond the scale $\overline{B}\sim B_0$. This unusual
 manifestation of  classical localization is also found in
non-interacting systems with a scalar random potential
\cite{Fogler97}. Finally, it is worth noting that, however
sophisticated the dissipative conductivity network may be, the
macroscopic Hall conductivity $\sigma_{xy}$ in the metallic system
at $\overline{B}\gg B_0$ assumes the collisionless form
$\sigma_{xy}=2\pi n/\overline{B}=p\gg\sigma_{xx}$ \cite{Trugman83}.

Having found the monotonic component of $\sigma_{xx}(\overline{B})$,
we turn to the magnetooscillations. These are conventionally
interpreted as the SdH effect for the CFs. We want to compute the
disorder-induced Dingle factor of the SdH oscillations
$\delta\sigma^{SdH}_{xx}\propto \cos (4\pi^2n/\overline{B})\exp
[-S_r(\overline{B})]$. At $\overline{B}\gg B_0$ the damping exponent
is given by $S_r={1\over 2}\left<\Phi^2\right>$, where $\Phi$ is the
random-$B$ flux through the cyclotron orbit in the field $\overline B$
\cite{Aronov95}. Had the CF dynamics been fully chaotic,
$\left<\ldots\right>$ here would have meant a disorder average over
all possible configurations of $B({\bf r})$, which yields
$S_r=4\pi^2p^4/(k_Fd)^2$ \cite{Aronov95}. The result might look
puzzling, since this $S_r$ is numerically by far larger
\cite{Aronov95} than what is observed experimentally. However, in the
limit $\overline{B}\gg B_0$, only the particles that drift along the
percolating trajectories with zero $B({\bf r})$ contribute to
$\sigma_{xx}$. Accordingly, whereas the above $S_r$ describes the
damping of the total density of states, the averaging we need to do
should be performed along the contours $B({\bf r})=0$. The result is
\begin{equation} S_r=(21\pi^2/ 4)p^8/(k_Fd)^6~. \end{equation} Both
the extremely sharp falloff of the amplitude of the oscillations with
increasing $p$ and the numerical value of $S_r$ are in fair agreement
with the experimental data available to date \cite{Du94}; yet, there
are good reasons to question this picture. The point is that the CF
system, like a conventional metal, will exhibit magnetooscillations
even if those of the density of states are totally
neglected. Specifically, in the diffusive regime, the quantum
interference of scattered waves leads to the QH oscillations
\cite{Pruisken90} $\delta\sigma^{QH}_{xx}\propto\cos
(2\pi\sigma_{xy})\exp [-S_l(\overline{B})]$, where \begin{equation}
S_l=2\pi\sigma_{xx}~. \end{equation} The confusion that one might have
at this point is over the notion that the QH effect transforms into
the SdH oscillations with decreasing magnetic field, which is
manifestly not true. Likewise, let us stress that the observation of
the QH oscillations at $p\gg 1$ does not require exponentially low
temperatures, -- while the characteristic localization length does
grow exponentially fast with $p$, the oscillations come from the
interference on the small scale of the effective mean free path. Note
that the periods of the QH and SdH oscillations coincide at
$\overline{B}\gg B_0$, but the damping factors are very different. The
quasiclassical approach which led us to the exponential falloff of
$\sigma_{xx}$ with decreasing $p$ fails and the QH effect shows up at
$p^{QH}\sim k_Fd/\ln^{1/2}(k_Fd)$, where $\sigma_{xx}$ drops to
unity. At $p\alt p^{QH}$ the oscillations are due to the QH effect.
This picture is in good agreement with experiment: from the typical
$\rho_{xx}\simeq 0.02\div 0.03$ at $p\simeq 6\div 7$, where the first
resistivity minimum is observed, one finds $\sigma_{xx}\simeq
\rho_{xx}p^2\simeq 1\div 1.5$. Furthermore, at smaller $p\leq 5$, the
values of $\sigma_{xx}$ at the maxima are close to ${1\over 2}$, thus
confirming that the CFs are indeed in the QH regime. As for the SdH
oscillations, Eq.\ (11) tells us that they might only be observable at
$p\alt p^{SdH}\sim (k_Fd)^{3/4}\ll p^{QH}$. The last inequality means
that in the limit $k_Fd\gg 1$ there is no room for the SdH effect. In
practical terms, however, $p^{QH}$ and $p^{SdH}$ are numerically close
at the experimentally relevant $k_Fd\sim 10\div 15$, so that the SdH
effect still can contribute to the first couple of minima ($p=7,6$).

In conclusion, we have studied the fermion kinetics in a smoothly
varying RMF with mean $\overline B$. We calculated the conductivity in
the regime of strong RMF and zero $\overline B$, when the transport is
determined by percolating snake-states. We demonstrated that
increasing $\overline{B}$ leads to the classical localization of
fermions, the key signature of which is the exponentially sharp drop
in $\sigma_{xx}$. We showed that this yields a strong enhancement of
the magnetooscillations, thus explaining why they are observed at
large $p$, where the SdH oscillations are negligible. We analyzed the
interplay between the SdH and QH oscillations and argued that the
latter are dominant in the limit $k_F d \gg 1$.

We are grateful to I. Aleiner, B. Altshuler, J. Hajdu, L. Rokhinson,
A. Shelankov, H. Stormer, and D. Tsui for interesting discussions.
This work was supported by the Deutsche Forschungsgemeinschaft (in
part through SFB195), by the German-Israeli Foundation, and by the
Russian Foundation for Basic Research under Grant No.\ 96-02-17894a.

\end{multicols}

\begin{references} 

\bibitem[*]{byline}Also at St.Petersburg Nuclear Physics Institute,
188350 St.Petersburg, Russia.

\bibitem[\dagger]{byline}Also at A.F.Ioffe Physico-Technical
Institute, 194021 St.Petersburg, Russia.

\bibitem{Jain89} D. K. Jain, Phys.\ Rev.\ Lett.\ {\bf 63}, 199 (1989);
Phys.\ Rev.\ B {\bf 40}, 8079 (1989); {\bf 41}, 7653 (1990).

\bibitem{Halperin93} B. I. Halperin, P. A. Lee, and N. Read, Phys.\
Rev.\ B {\bf 47}, 7312 (1993).

\bibitem{Lopez91} A. Lopez and E. Fradkin, Phys.\ Rev.\ B {\bf 44},
5246 (1991); {\bf 47}, 7080 (1993).

\bibitem{Willett95} R. L. Willett, K. W. West, and L. N. Pfeiffer,
Phys.\ Rev.\ Lett.\ {\bf 75}, 2988 (1995). 

\bibitem{Du94} R. R. Du {\it el al.,} Solid State Commun.\ {\bf 90},
71 (1994); Phys.\ Rev.\ Lett.\ {\bf 70}, 2944 (1993); D. R. Leadley
{\it et al.,} Phys.\ Rev.\ Lett.\ {\bf 72}, 1906 (1994); Phys.\ Rev.\
B {\bf 53}, 2057 (1996); P. T. Coleridge {\it et al.,} Phys.\ Rev.\ B
{\bf 52}, R11603 (1995); L. P. Rokhinson and V. J. Goldman, Report
No.\ cond-mat/9704043.

\bibitem{Aronov95} A. G. Aronov, E. Altshuler, A. D. Mirlin, and
P. W\"olfle, Phys.\ Rev.\ B {\bf 52}, 4708 (1995); A. D. Mirlin,
E. Altshuler, and P. W\"olfle, Ann.\ Phys.\ (Leipzig) {\bf 5}, 281
(1996).

\bibitem{Khveshchenko96} D. V. Khveshchenko, Phys.\ Rev.\ Lett.\ {\bf
77}, 1817 (1996).

\bibitem{Chklovskii93} D. B. Chklovskii and P. A. Lee, Phys.\ Rev.\ B
{\bf 48}, 18060 (1993); D. K. K. Lee, J. T. Chalker, and D. Y. K. Ko,
Phys.\ Rev.\ B {\bf 50}, 5272 (1995).

\bibitem{Isichenko92} M. B. Isichenko, Rev.\ Mod.\ Phys.\ {\bf 64},
961 (1992).

\bibitem{Fogler97} M. M. Fogler, A. Yu.\ Dobin, V. I. Perel, and
B. I. Shklovskii, Report No.\ cond-mat/9702121.

\bibitem{Trugman83} S. A. Trugman, Phys.\ Rev.\ B {\bf 27}, 7539
(1983).

\bibitem{Pruisken90} A. M. M. Pruisken, in {\it The Quantum Hall
Effect}, edited by R. E. Prange and S. M. Girvin (Springer, Berlin,
1990).

\end{references}
\end{document}